\documentclass[amsmath,amssymb,nofootinbib,prd]{revtex4}

\usepackage{amsmath}
\usepackage{bm}
\usepackage{color}
\usepackage{epsfig}
\usepackage{mathrsfs}

\renewcommand{\mathbf}{\bm}
\newcommand{\be}{\begin{equation}}
\newcommand{\ee}{\end{equation}}
\newcommand{\ba}{\begin{eqnarray}}
\newcommand{\ea}{\end{eqnarray}}
\newcommand{\bei}{\begin{itemize}}
\newcommand{\eei}{\end{itemize}}

\newcommand{\omegam}{\Omega_{{m}}}
\newcommand{\omegab}{\Omega_{{b}}}
\newcommand{\omegac}{\Omega_{{c}}}
\newcommand{\omegal}{\Omega_{{\Lambda}}}
\newcommand{\omegamo}{{\Omega_{m}}}
\newcommand{\omegmo}{{\Omega_{m}}}

\newcommand{\omegko}{{\Omega_{K}}}
\newcommand{\omoin}{{\Omega^{\textnormal{in}}_{m}}}
\newcommand{\omoout}{{\Omega^{\textnormal{out}}_{m}}}
\newcommand{\kapr}{\kappa(r)r^2}
\newcommand{\Hperp}{H_{\perp}}
\newcommand{\Hperpo}{H_{\perp 0}}

\newcommand{\Hpar}{H_{\parallel}}
\newcommand{\aperp}{a_{\perp }}

\newcommand{\apar}{a_{\parallel}}
\newcommand{\dA}{d_{\textnormal{A}}}
\newcommand{\LCDM}{\Lambda\textnormal{CDM}}
\newcommand{\xvec}{\mathbf{r}}
\newcommand{\kvec}{\mathbf{k}}
\newcommand{\xhat}{\hat{\mathbf{r}}}
\newcommand{\khat}{\hat{\mathbf{k}}}
\newcommand{\xiDel}{\xi}
\newcommand{\xiDelperp}{\xi^{\perp}}
\newcommand{\xiDelpar}{\xi^{\parallel}}

\newcommand{\ulm}{_{\ell m}}

\newcommand{\ulmp}{_{\ell' m'}}
\newcommand{\Ylm}{Y_{\ell m}}
\newcommand{\Ylmp}{Y_{\ell' m'}}
\newcommand{\opL}{\mathcal{L}}

\newcommand{\Hoin}{H^{\textnormal{in}}_0}

\begin{document}

\title{Galaxy correlations and the BAO in a void universe:\\ structure formation as a test of the Copernican Principle}

\date{\today}

\author{Sean February$^1$}
\author{Chris Clarkson$^1$}
\author{Roy Maartens$^{2,3}$}
\affiliation{ $^1\,$Astrophysics, Cosmology \& Gravity Centre,
Department of Mathematics and Applied Mathematics, University of
Cape Town, Cape Town 7701, South Africa\\
$^2\,$Department of Physics, University of Western Cape, Cape Town
7535, South Africa\\
$^3\,$Institute of Cosmology \& Gravitation, University of
Portsmouth, Portsmouth PO1 3FX, UK}

\begin{abstract}

A suggested solution to the dark energy problem is the void model, where accelerated expansion is replaced by Hubble-scale inhomogeneity. In these models, density perturbations grow on a radially inhomogeneous background. This large scale inhomogeneity distorts the spherical Baryon Acoustic Oscillation feature into an ellipsoid which implies that the bump in the galaxy correlation function occurs at different scales in the radial and transverse correlation functions. We compute these for the first time, under the approximation that curvature gradients do not couple the scalar modes to vector and tensor modes.  The radial and transverse correlation functions are very different from those of the concordance model, even when the models have the same average BAO scale. This implies that if void models are fine-tuned to satisfy average BAO data, there is enough extra information in the correlation functions to distinguish a void model from the concordance model. We expect these new features to remain when the full perturbation equations are solved, which means that the radial and transverse galaxy correlation functions can be used as a powerful test of the Copernican Principle.

\end{abstract}

\maketitle

\section{Introduction}
\label{sec:intro}

The homogeneous and isotropic
Friedmann-Lema\^\i tre-Robertson-Walker (FLRW) universe is based on the assumptions
that the Copernican Principle holds true and that the universe is isotropic about
our location. The high degree of isotropy of the Cosmic Microwave Background (CMB) provides strong support for the second assumption. The Copernican assumption cannot be directly tested, although null tests of this assumption have been devised (for reviews, see \cite{Clarkson:2010uz,Clarkson:2012bg}). A spatially flat FLRW universe containing 20\% cold dark
matter (CDM), 5\% baryons and 75\% dark energy in the form of the cosmological
constant ($\Lambda$) provides an excellent fit to the wide range of observations to date. However, there is still no satisfactory theoretical
explanation from fundamental physics for the observed value of $\Lambda$. This has prompted many authors to consider
alternatives to $\LCDM$, such as modified gravity \cite{Clifton:2011jh}, 
backreaction \cite{Clarkson:2011zq}, and spherically symmetric but inhomogeneous exact solutions
to Einstein's equations known as Lema\^itre-Tolman-Bondi (LTB) models (see \cite{Marra:2011ct,Clarkson:2012bg} for recent reviews).

In this work we focus on the latter alternative. Hubble-sized LTB void models break the Copernican Principle by
placing our galaxy in a special position, at the centre of an underdense region of $O$(Gpc) scale. The simplest of these models are able to fit the distance-redshift data from type Ia supernovae (SNIa) and the
CMB, though this gives considerable tension with the locally measured value of $H_0$ (see e.g.,~\cite{Biswas:2010xm}). Attempts to overcome this require the relaxation of assumptions on the homogeneity of the early universe in the form of isocurvature modes~\cite{Clarkson:2010ej} or a change to the primordial power spectrum~\cite{Nadathur:2010zm}. 

It remains an open issue 
whether they can fit the Baryon Acoustic Oscillation (BAO) data, and the more general data on the growth and distribution of large scale structure. Previous papers \cite{GarciaBellido:2008nz, Zibin:2008vk,GarciaBellido:2008yq, Biswas:2010xm, Zumalacarregui:2012pq}
have computed the BAO scales in a geometric approximation, using  the anisotropic
expansion rates of the background model but ignoring any effects from the anisotropic
growth of structure in void models. This is not surprising, since structure formation on an LTB background has not yet been calculated, because it is much more complicated than in the standard model~\cite{Clarkson:2007yp,Zibin:2008vj, Clarkson:2009sc, Dunsby:2010ts, Alonso:2010zv,Nishikawa:2012we}.  

Here we calculate for the first time the 2-point correlation function on an LTB background, and use it to extract the radial and transverse BAO scales. This incorporates the effects of the evolution of density perturbations on an LTB background, using the perturbation formalism developed in \cite{Clarkson:2009sc}. We neglect the coupling of scalar to vector and tensor modes in the metric potentials. This is expected to be a good approximation for the simplest LTB models, in which the background shear $-$ responsible for the coupling of the modes in the first place $-$ is typically of the order of a few percent~\cite{GarciaBellido:2008yq}. (The accuracy of this approximation is under investigation via numerical solutions \cite{full_system_check}.)

LTB models have enough freedom to always fit the average BAO scale. In these models the proper radius of the sound horizon at the drag epoch is approximately given by~\cite{Clarkson:2010ej,Clarkson:2012bg} 
\be \label{ds}
d_s=\frac{121.4\ln\left({2690f_b}/{\eta_{10}}\right)}{\sqrt{1+0.149{\eta_{10}}^{3/4}}}\left[\frac{1\,\text{K}}{T_d(f_b,\eta_{10})}\right]\,\text{Mpc}\,,
\ee
where $N_\text{eff}=3.04$,
$f_b=\Omega_b/\Omega_m$ is the local baryon fraction, $\eta=10^{-10}\eta_{10}$ is the baryon-photon ratio at that time,  $T_d$ is the temperature at the drag epoch and it is assumed that during the process of recombination, the scale of the void inhomogeneity is much larger than the horizon size at that time ($\sim 100\,$Mpc).
In general, $f_b$ and $\eta_{10}$ have radial degrees of freedom in them, and are no longer measured by the CMB at the radial scales of interest for the BAO. Consequently, $d_s$ is not constant spatially, and can vary over the scale of the model. Therefore, measuring the mean BAO scale in some shell around us cannot place constraints on late-time inhomogeneity without some other measurement of $f_b$ and $\eta$ in the same shell at early times~-- which lie inside our past lightcone. In fact, the models can even be fine-tuned to have the same radial and angular BAO scales, by altering the bang time function appropriately. 

As we shall show, however, the radial and transverse 2-point correlation functions contain much more information than the peak positions which determine the BAO scales. In real space, the radial and transverse correlation functions are typically very different from each other. To make them close to those of the $\LCDM$ model with the same BAO scale would require high levels of fine-tuning in either $d_s$ or the primordial power spectrum. Effectively, the radial and transverse 2-point correlation functions can thus be used as an important probe of the Copernican assumption.

This paper is organized as follows. In Section \ref{sec:bao} we recap the standard geometric approximation for computing BAO scales and describe our method for determining the background dynamics of the LTB spacetime. In Section \ref{sec:perts} we provide an overview of perturbation theory in LTB via the 2+2 decomposition approach. This is followed by a derivation of the anisotropic two-point correlation function for the gauge-invariant matter density perturbation in Section \ref{sec:2pcf}, where we present the computation of the correlations and the BAO scales. Finally, in Section \ref{sec:discuss} we discuss the consequences of our results.

\section{Background model and evolution of the BAO scales}
\label{sec:bao}

The background void model is described by the LTB metric,
 \be \label{LTBmetric2}
ds^2 = -dt^2 + \frac{\apar^2(t,r)}{1-\kapr}dr^2 +
\aperp^2(t,r)r^2d\Omega^2\,,~~\apar = {\dA}' \,,~ \dA=r\aperp\,,
 \ee
where $\dA$ is the angular diameter distance, and a prime indicates $\partial/\partial r$. In the FLRW limit, $\apar(t,r)=\aperp(t,r)=a(t)$ and $\kappa(r)=K$. The expansion rates
transverse to and along the line-of-sight are
 \be\label{H}
\Hperp(t,r) = \frac{\dot a_\perp(t,r)}{\aperp(t,r)}~~\text{and}~~\Hpar(t,r) =
\frac{\dot a_{\|}(t,r)}{\apar(t,r)}\,.
 \ee 
The past lightcone
of the central observer has null geodesics that are given by
 \be
\frac{dt}{dz} = -\frac{1}{(1+z)\Hpar(t(z), r(z))}\,, ~~
\frac{dr}{dz} = \frac{\left[1-\kappa(r(z))
r^2(z)\right]^{1/2}}{(1+z)\apar(t(z), r(z))\Hpar(t(z), r(z))}\,.
 \ee
We use the notation $F(z) \equiv F(t(z),r(z))$ to denote evaluation on the past lightcone.
 
The anisotropic expansion rates (\ref{H}) act on the acoustic sphere of proper radius $L_*$ at an initial high redshift $z_*$, so that by redshift $z$ it has evolved into an
axisymmetric ellipsoid with semi-axes 
\cite{Clarkson:2007pz,Zibin:2008vk}
 \ba
 \label{lexppar}
L^\text{geo}_\parallel(z) = L_* \frac{\apar(z)}{\apar(t_{*},r(z))},~~~
L^\text{geo}_\perp(z) = L_*\frac{\aperp(z)}{\aperp(t_{*},r(z))}.
 \ea
However, this geometric approximation does {\em not} give the correct BAO feature in the galaxy distribution~-- because it neglects the anisotropic effects of perturbations in LTB and their impact on the correlation function. Previous work \cite{Zibin:2008vk,Biswas:2010xm,Zumalacarregui:2012pq,
GarciaBellido:2008yq,Moss:2010jx,Nadathur:2010zm} on comparing the BAO scales in LTB with observations has all neglected the effects of LTB perturbations.
Below we fill this gap by computing the correlation functions associated with the density perturbation and then extracting the BAO scales from the correlation functions.
 
The observable quantities of the BAO feature are its redshift
extent $\delta z(z)$ and angular size $\delta \theta(z)$. These are converted to the physical radial and transverse length scales via
 \ba
 \label{lphys}
L_\parallel(z) ={\delta z(z) \over (1+z)\Hpar(z)}, ~~~
L_\perp =\dA(z)\, \delta\theta(z),
 \ea
for small $\delta z$ and $\delta\theta$. Note that we neglect redshift space distortions.

The quantity \cite{Biswas:2010xm}
 \be \label{dzbao}
d_z = \left[ {(\delta\theta)^2 \delta z \over z}\right]^{1/3},
 \ee
encodes an average of the two observable scales of the sound ellipsoid. In an
FLRW model it reduces to
\ba
d_z = {L_*(1+z_*) \over D_V}, ~~~ D_V(z) = \left[(1+z)^2\dA^2(z)\frac{z}{H(z)}\right]^{1/3},
\ea
where $L_*(1+z_*)$ is the comoving sound horizon and $D_V$ is the standard volume-averaged BAO scale \cite{Eisenstein:2005su}.

The LTB analogue of the Friedmann equation is
 \be
\frac{\Hperp^2(t,r)}{\Hperpo^2(r)}={\omegmo(r) \over \aperp^3(t,r)} + {\omegko(r)
\over \aperp^2(t,r)}, ~~~\mbox{where}~~~ \omegmo(r)+ \omegko(r)=1,~~a_\perp(t_0,r)=1,~~\Hperpo(r)=\Hperp(t_0,r).
 \ee
The
observed Hubble constant is  $\Hoin\equiv\Hperpo(0)=100h$ km/s/Mpc, where ``in''
indicates evaluation at the centre. For open LTB models, the
parametric solution is
 \ba
\aperp (t,r) &=& \frac{\omegamo(r)[\cosh{2u(t,r)}-1]}
{2\left[1-\omegamo(r)\right]} \,, \label{aperp}\\
t&=& \frac{\omegamo(r)[\sinh{2u(t,r)} -
2u(t,r)]}{2\Hperpo(r)\left[1-\omegamo(r)\right]^{3/2}}\,,
\label{time}
 \ea
where we choose a simultaneous big bang (uniform bang time function). Setting $t=t_0$ in
(\ref{aperp}) and (\ref{time}) gives
 \ba
u_0(r) = \frac{1}{2}\cosh^{-1}\left[{2\over
\omegamo(r)}-1\right],~~ \Hperpo(r) =
\frac{\omegamo(r)[\sinh{2u_0(r)} -
2u_0(r)]}{2t_0\left[1-\omegamo(r)\right]^{3/2}}.
 \ea
Thus $\Hperpo(r)$ is determined when $\omegamo(r)$ and $t_0$ are
chosen. Then (\ref{time}) determines $u(t,r)$ and
$\aperp(t,r)$ follows from (\ref{aperp}).

For the purposes of this study, we choose a simple Gaussian void profile for the dimensionless density parameter,
 \ba \label{void}
\omegamo(r) = \omoout - (\omoout-\omoin)\exp\left( -\frac{r^2}{\sigma^2} \right),~~\mbox{with}~ \omoout=1, ~ h=0.7\,,
 \ea 
where ``out" refers to the asymptotic
Einstein-de Sitter region, and $\sigma$ characterizes the size of the void (see \cite{February:2009pv} for details). The physical matter density is then 
 \ba
\rho(t,r) = \frac{\omegamo(r)\Hperpo^2(r)}{8\pi G \apar(t,r)\aperp^2(t,r)}\left[3 + r\left\{2\frac{\Hperpo
'(r)}{\Hperpo(r)} + \frac{\omegamo'(r)}{\omegamo(r)} \right\}\right] .
 \ea

%
\begin{figure}[htpb]
\centering \epsfig{file=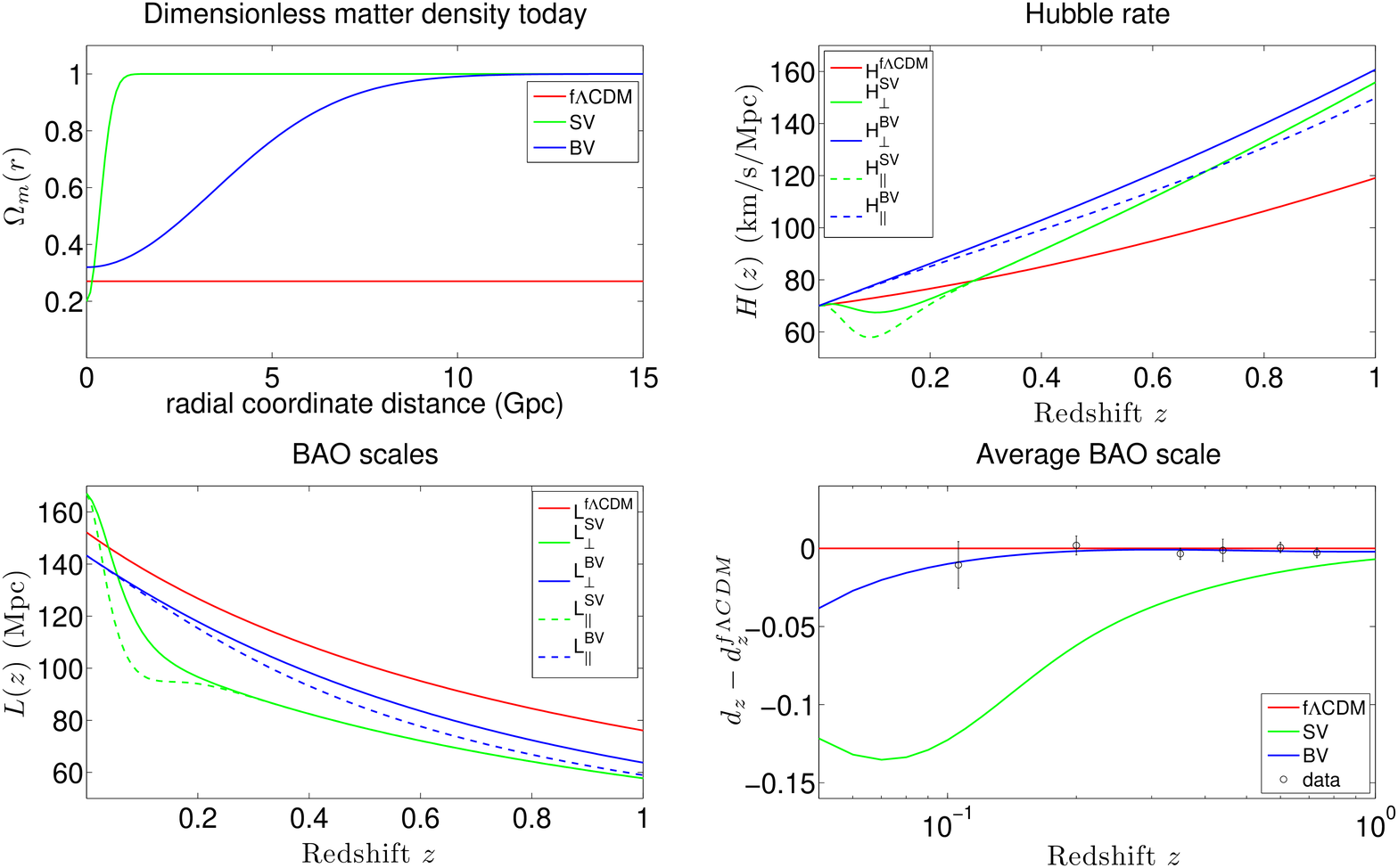,width=1.05\textwidth}
 \caption{{\em Upper:} Background density and expansion rates for the models \eqref{concpar}--\eqref{bv}. {\em Lower:} Using the geometric approximation \eqref{lexppar}, the evolution of the BAO length scales  (left), and the average BAO scale (right). Black circles indicate
 measurements from \cite{Blake:2011en}. }\label{bgscales}
\end{figure}

\section{Scalar Perturbations on an LTB background}
\label{sec:perts}

The full perturbation theory on an LTB background is developed in
\cite{Clarkson:2009sc} via a 2+2 split of the spacetime, which makes explicit the coupling of vector and tensor modes to scalar modes at linear order.   A first approximation is to neglect this mode-mixing, and focus only on `scalar' modes which occur in the even parity sector. Then the
perturbed metric in Regge-Wheeler gauge is (\cite{Clarkson:2009sc}, with notational change, $ \varphi \to -2\Phi$)
 \be
 \label{pert_metric_LTB}
ds^2=- \left[1+2\Phi(t,\mathbf{x})\right]dt^2 + \left[1-2\Phi(t,\mathbf{x})\right]\bar{g}_{ij}dx^idx^j\,,
 \ee 
where $\bar{g}_{ij}$ is the spatial part of \eqref{LTBmetric2}. The Newtonian potential 
obeys a simple generalization, without gradients, of the FLRW evolution equation for the Newtonian potential \cite{Clarkson:2009sc}:
 \ba \label{phidot}
\ddot{\Phi} + 4\Hperp\dot{\Phi} - \frac{2\kappa}{\aperp^2}\Phi = 0\,.
 \ea 
Because there are no spatial gradients, $\Phi$ evolves independently in each $r=\,$const shell, as if in a separate dust FLRW model. This does not mean that there is no dependence on spatial gradients: density fluctuations depend on spatial gradients of $\Phi$ which couple to the anisotropic expansion of the model. 
The gauge-invariant matter density perturbation $\Delta$ is found via the equivalent of the Poisson equation in LTB~\cite{Clarkson:2009sc}:
 \ba
 \label{poiss_LTB}
  4\pi G \apar^2\rho \Delta  &=& \mathcal{L} \left[ \Phi\right] , \\
\text{where}~~~  \mathcal{L} &=& (1-\kappa r^2)\partial_r^2 + \left[\frac{2\apar}{\aperp r} - \left(1+\frac{2\apar}{\aperp}\right)\kappa r  - \frac{r^2\kappa'}{2} - \frac{{\apar}'}{\apar}\left(1-\kappa r^2\right) \right] \partial_r \nonumber \\
 &&{}  -{\apar^2 \over \aperp^2}\frac{\ell(\ell+1)}{ r^2}  + {\apar \over \aperp}\left[ r\kappa' + \left(2 +{\apar \over \aperp}\right) \kappa \right] -\apar^2(\Hpar+2\Hperp) \partial_t -\aperp^2 \Hperp(\Hperp+2\Hpar). \label{lop}
 \ea
(Recall that the LTB model contains only CDM and baryons.) In FLRW, we recover the standard Poisson equation:
 \ba
4\pi G a^2\rho \Delta &=& \left[\vec\nabla^2+ 3K \right] \Phi - 3a^2H (\dot{\Phi} + H\Phi),
\label{poiss_FL}\\\text{where}~~~ \vec{\nabla}^2 &=& (1-K r^2)
\partial_r^2 + \frac{(2-3K r^2)}{r}\partial_r -
\frac{\ell(\ell+1)}{r^2} \,.
 \ea
Here $\ell$ is the angular wave number in a spherical harmonic expansion,
 \ba
\Phi(t,\mathbf{x}) = \sum_{\ell m}\Phi_{\ell m}(t,r) Y_{\ell m}(\theta,\varphi), 
 \ea 
and similarly for $\Delta$.

 \begin{figure}[htbp]
\begin{center}
\centering \epsfig{file=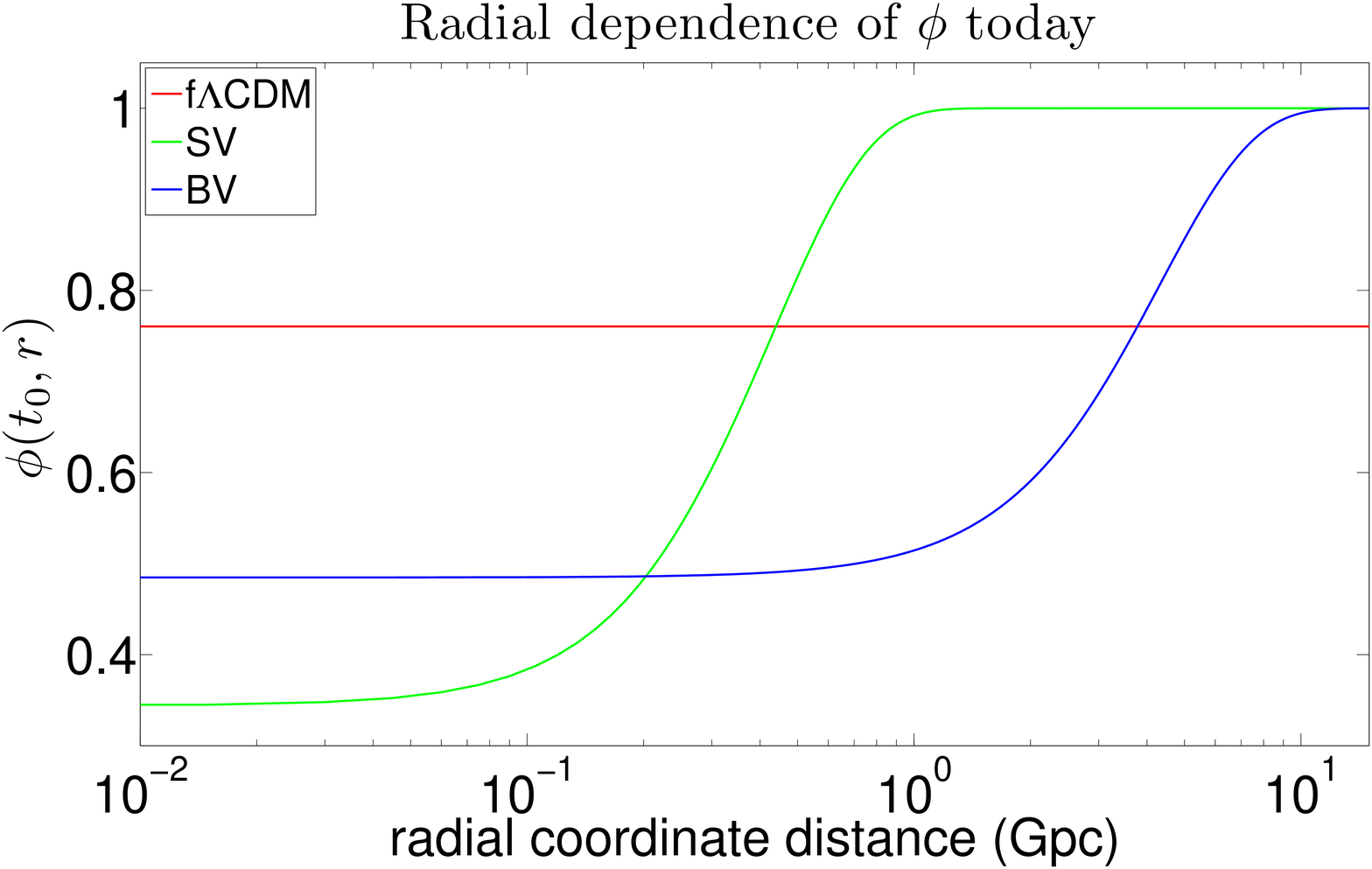,width=0.49\textwidth}
\epsfig{file=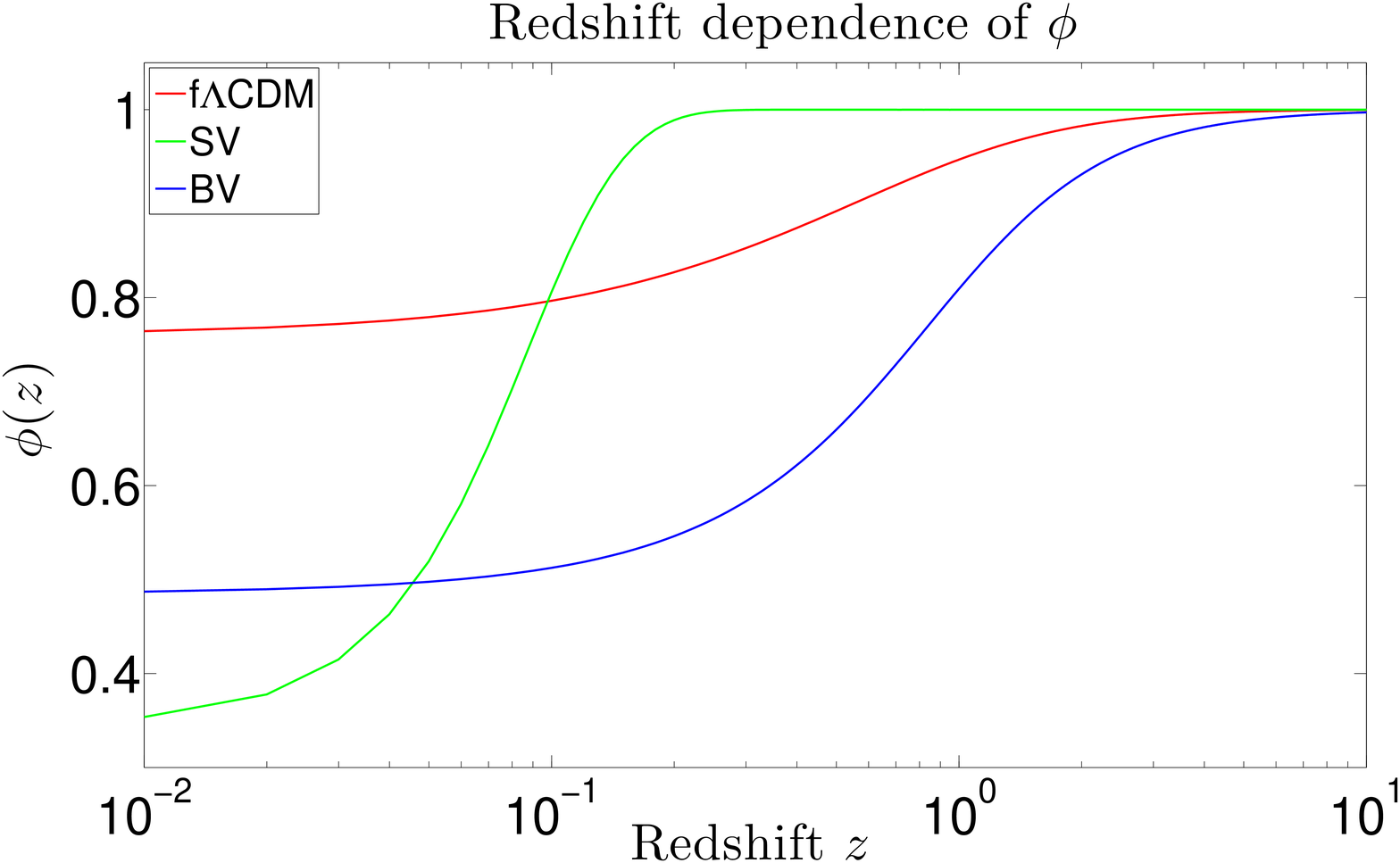,width=0.49\textwidth}
\caption{The gravitational potential $\phi$ as a function of radius  today (left), and of redshift (right). }
\label{fig:phi}
\end{center}
\end{figure}
 \begin{figure}[htbp]
\begin{center}
\centering 
\epsfig{file=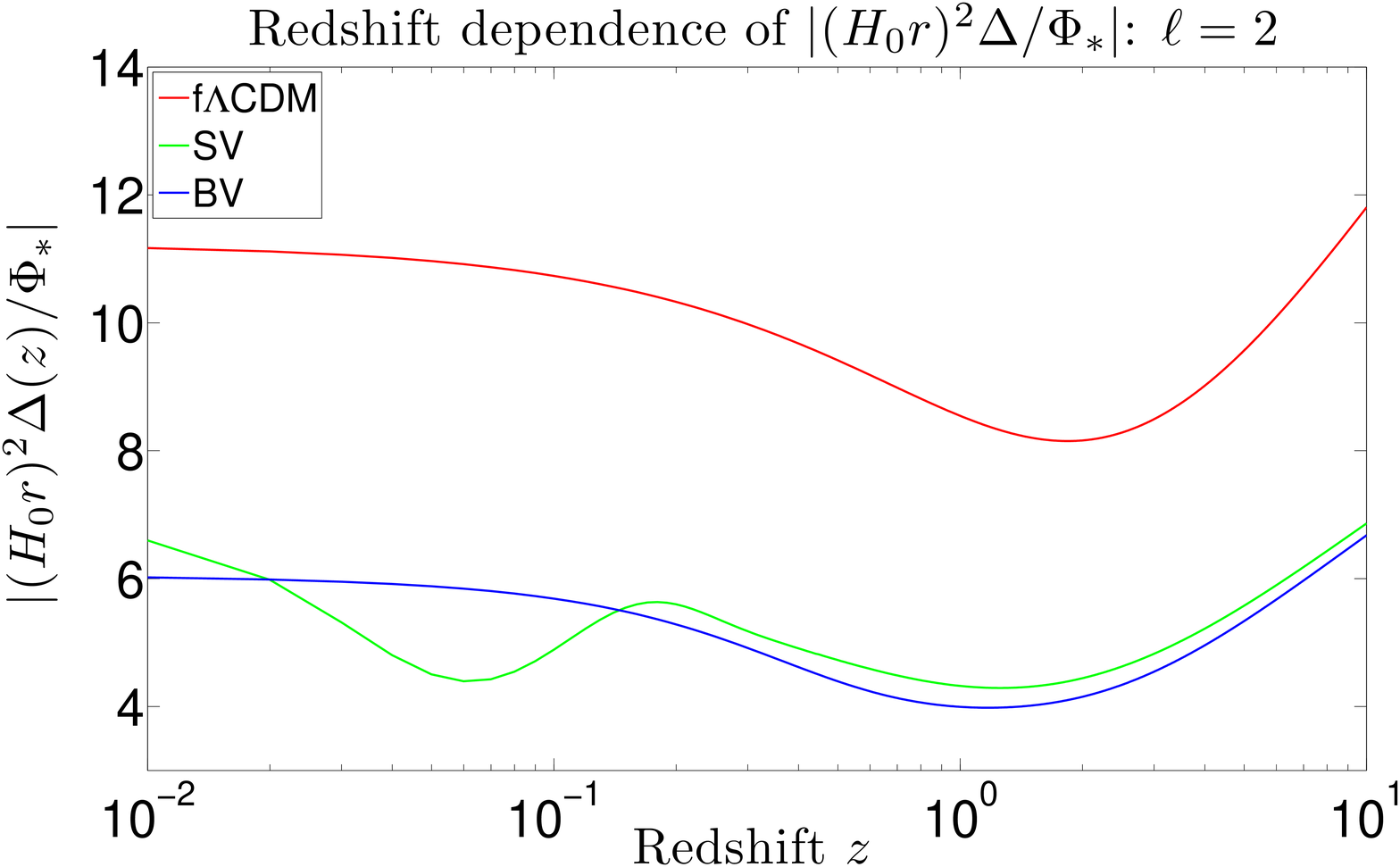,width=0.49\textwidth}
\epsfig{file=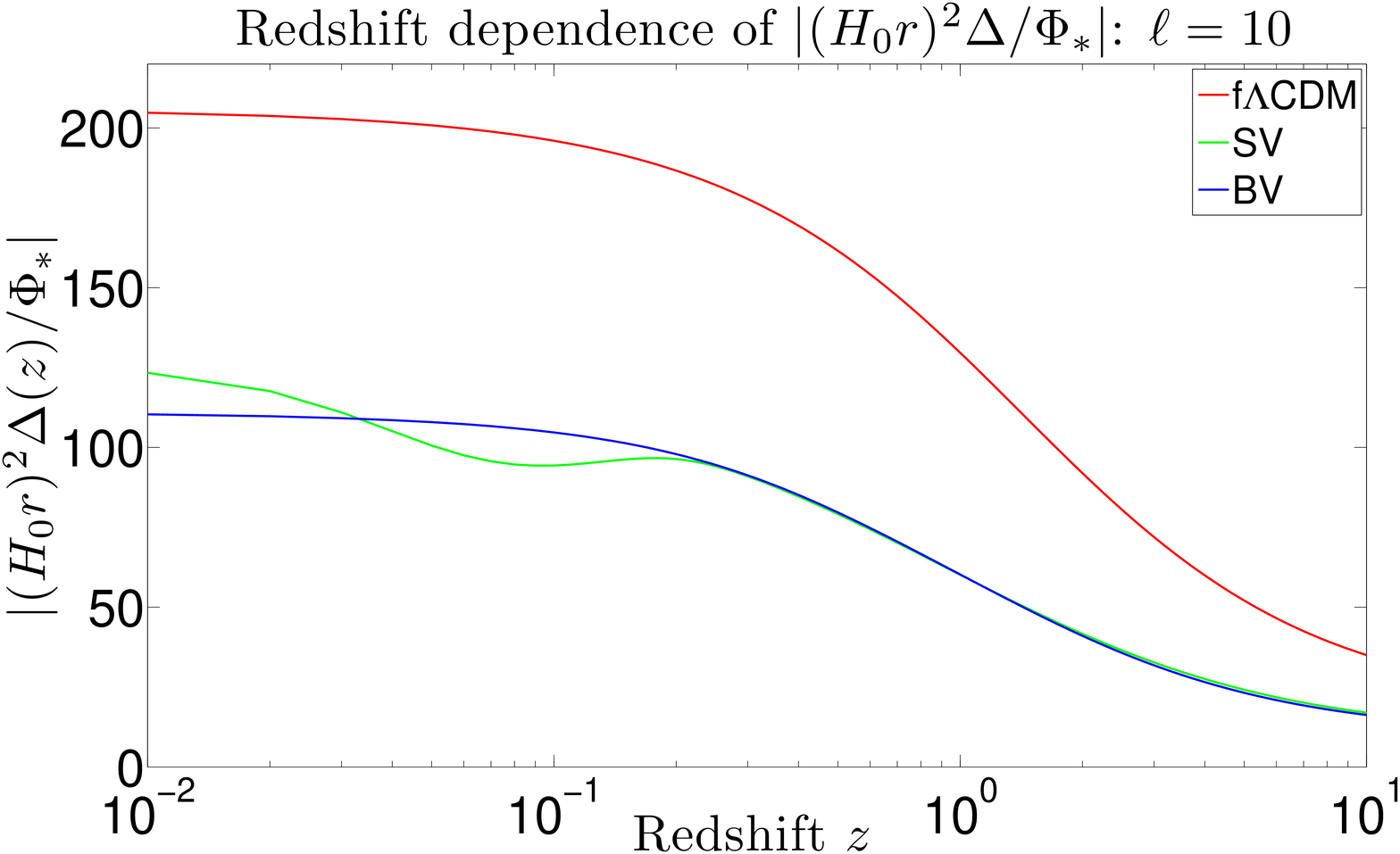,width=0.49\textwidth}
\caption{The normalized density perturbation $|(H_0r)^2\Delta_{\ell m}/\Phi_{*\ell m}|$ against redshift, for small (left) and large (right) $\ell$. }
\label{fig:delta}
\end{center}
\end{figure}

We set initial conditions for $\Phi$ at a high redshift, $z_*=100$,
where we assume the background is effectively FLRW. We write
 \be
\Phi_{\ell m}(t,r)=\phi(t,r)\Phi_{*\ell m}(r),~~ \phi(t_*,r)=1\,.
 \ee
The subsequent time
evolution of $\phi(t,r)$ is then determined by (\ref{phidot}) for each $r$. Using \eqref{aperp} and \eqref{time}, this implies
 \ba
\phi(t,r) &=& C(r) \frac{\cosh{u(t,r)}}{\sinh^5{u(t,r)}}\big[\sinh{2u(t,r)} -6u(t,r) + 4\tanh{u(t,r)}\big]\,, \\
C(r) &=& \frac{\sinh^5{u_*{}(r)}}{\cosh{u_*{}(r)} \big[\sinh{2u_*{}(r)}-6u_*{}(r)+4\tanh{u_*{}(r)}\big]}.
 \ea
 Now, $\Phi_{*\ell m}(r)$ can be written as
 \be
\label{philm_init}
\Phi_*{}_{\ell m}(r) = \sqrt{\frac{2}{\pi}}\,i^{\ell}\int d^3k \,
j_{\ell}(kr) \Phi_*{}(\kvec) Y_{\ell m}(\khat)\,,
\ee
which is related to the power spectrum via
\be
\left\langle \Phi_*{}\left(\kvec_1\right)  \Phi_*{}\left(\kvec_2\right) \right\rangle =\frac{2\pi^2}{k_1^3} \mathcal{P}_{\Phi}{}_*\left( k_1\right) \delta^3\left( \kvec_1 + \kvec_2\right). \label{pphi}
 \ee
The initial power spectrum of the Newtonian potential is given by
 \ba
\mathcal{P}_{\Phi_*}(k) = \frac{9}{25} \mathcal{P}_{\mathcal{R}}(k_0) T^{2}(k)\,,
 \ea
where $\mathcal{P}_{\mathcal{R}}(k_0)=2.41\times 10^{-9}$ is the amplitude of the primordial curvature perturbation on the scale $k_0=0.002$ Mpc$^{-1}$, and $T(k)$
is the matter transfer function, with $T(k_0) \approx 1$. The concordance parameters \eqref{concpar} are used in the fitting formula of \cite{Eisenstein:1997ik} to compute $T(k)$, which is employed in all of the models.

Note that when using a flat FLRW initial power spectrum, we need to use
the flat FLRW comoving coordinate $r_F$ in (\ref{philm_init}) at $t_*$, as opposed to the LTB coordinate
$r$. Proper radial distance
is independent of coordinates: $d_p(t_*,r_F) =  d_p(t_*,r)$. Since $ d_p(t_*,r_F)=a(t_*)r_F$, we find that
 \ba
r_F =  (1+z_*) \int_0^{r} dr \,\frac{\apar(t_*,r)}{\sqrt{1-\kappa(r)r^2}}  \, \equiv \, f(r) \,,
 \ea 
where $f(r) \approx (1+z_*)\aperp(t_*,r) r$ since $\sqrt{1-\kappa(r)r^2} \approx 1$ for all $r$ and $\apar = \partial_r( \aperp r)$.  Then (\ref{philm_init}) becomes
 \ba
\Phi_{*\ell m}(r) = \sqrt{\frac{2}{\pi}}\,i^{\ell}\int d^3k \, j_{\ell}(kf(r)) \Phi_*(\kvec) \Ylm(\khat)\,.
 \ea

\section{correlation functions and the BAO scales}
\label{sec:2pcf}

The two-point correlation function (2PCF) for the density
perturbation $\Delta$, as observed by a central observer down their past lightcone, is 
 \ba 
\xiDel(t_1,t_2,\xvec_1,\xvec_2)& \equiv& \langle \Delta(t_1,\xvec_1)
\Delta(t_2,\xvec_2) \rangle= \xiDel(t(z_1),t(z_2),r(z_1)\xhat_1,r(z_2)\xhat_2) \nonumber\\ \label{xi_starting_point}
&=& \xiDel(t(z_1),t(z_2),r(z_1),r(z_2),\delta\theta),~~~\text{where}~ \xhat_1\cdot\xhat_2=\cos \delta\theta.
 \ea
The second line follows from statistical isotropy, which applies for central
observers.  We neglect redshift space distortions for simplicity,  since we are not testing the void models against data but only comparing them with the concordance model. We also neglect all complications from bias for the same reason.

Using the Poisson equation \eqref{poiss_LTB}, the correlation function 
(\ref{xi_starting_point}) becomes
 \ba
\xiDel(z_1,z_2,\delta\theta) &=& \left[(4\pi G {\apar}_1 {\apar}_2)^2\rho_1\rho_2\right]^{-1}\sum_{\ell m,\ell' m'}  \opL_1 \phi_1 \,\opL_2 \phi_2  Y\ulm(\xhat_1) Y\ulmp(\xhat_2)
\langle  {\Phi_{*1}}\ulm {\Phi_{*2}}\ulmp \rangle  \nonumber\\
&=&\left[8 \pi^3 (G {\apar}_1{\apar}_2)^2\rho_1\rho_2\right]^{-1}\sum_{\ell m,\ell' m'} i^{\ell-\ell'} \int d^3k_1 \, d^3k_2 \, \mathcal{L}_1\big[\phi_1j_{\ell}(k_1 f_1)\big] \mathcal{L}_2 \big[ \phi_2 j_{\ell'}(k_2f_2)  \big] Y\ulm(\xhat_1) Y\ulmp(\xhat_2) \nonumber \\
& &~\times  \left\langle \Phi_*{}(\kvec_1)  \Phi_*{}(\kvec_2) \right\rangle \Ylm(\khat_1) \Ylmp(\khat_2)\,,
\label{xiDel_convert_pk}
 \ea
 where a subscript $i=1,2$ on a function of $(t,r)$ means the quantity is evaluated at $(t(z_i),r(z_i))$.
Using \eqref{pphi} and standard identities in
(\ref{xiDel_convert_pk}), we get
 \ba
 \label{xi_LTB}
\xiDel (z_1,z_2,\delta\theta)
&=& \left[(4 \pi G {\apar}_1 {\apar}_2)^2 \rho_1\rho_2\right]^{-1} \sum_{\ell} {(2\ell+1) P_{\ell}(\cos{\delta\theta})} \int \frac{dk}{k} \, {{\cal J}_\ell}(z_1,k) {{\cal J}_\ell}(z_2,k) \mathcal{P}_{\Phi}(k) \,, \label{2pcf}\\ \text{where}~~~
{{\cal J}_\ell}(z,k)& =& {\cal L}\big[\phi(t(z),r(z))j_\ell (kf(z))\big] \label{bigjl}
 \ea 

To evaluate \eqref{bigjl}, we use \eqref{lop} and the following identities for the spherical Bessel function
 \ba
\label{diff_jl_r_final}
\partial_rj_{\ell}(kf) &=& \ell\frac{f'}{f} j_{\ell} - k f' j_{\ell+1} \,,\\
\partial^2_rj_{\ell}(kf) &=& \left[\ell \frac{f''}{f} + \ell(\ell-1)\frac{f^{\prime 2}}{f^2} - k^2 f^{\prime 2} \right] j_{\ell} - \left(f'' - 2\frac{f^{\prime 2}}{f}\right) k j_{\ell+1}\,.
 \ea
The result is
 \ba
{{\cal J}_\ell} &=& \left[\alpha + \beta \ell + \gamma \ell^2 - (1-\kappa r^2) f^{\prime 2} k^2 \phi \right]  j_{\ell} - \nu k j_{\ell+1}\,,
\label{curlyj_LTB_1}
 \ea
where
 \ba
\alpha &=& (1-\kappa r^2)\phi'' + A\phi' -\apar^2(\Hpar+2\Hperp)\dot{\phi} + B\phi ,\\
\beta &=&  \left[(1-\kappa r^2)\left(\frac{f''}{f}-\frac{f^{\prime 2}}{f^2}\right)+A{f' \over f} - \frac{\apar^2}{r^2 \aperp^2} \right]\phi  +  2(1-\kappa r^2)
{f' \over f}\phi' ,\\
\gamma &=&  \left[(1-\kappa r^2){f^{\prime 2} \over f^2} - \frac{\apar^2}{r^2 \aperp^2}  \right] \phi\,, \\
\nu &=& \left[ (1-\kappa r^2)\left(f'' - \frac{2f^{\prime 2}}{f}\right)+Af'\right]\phi + 2f'(1-\kappa r^2)\phi'\,,
 \ea 
 and
 \ba
A &=& \frac{2\apar}{\aperp r} - \left(1+\frac{2\apar}{\aperp}\right)\kappa r  - \frac{r^2\kappa'}{2} - \frac{{\apar}'}{\apar}\left(1-\kappa r^2\right)  \,,\\
B &=&  -\apar^2 \Hperp(\Hperp+2\Hpar) +{\apar \over \aperp} {\left[r\kappa' + \left(2+\frac{\apar}{\aperp}\right) \kappa \right]}. \label{term_B}
 \ea
In the flat FLRW case, (\ref{curlyj_LTB_1}) becomes
\ba
{{\cal J}_\ell}(z,k) &=& -\Big[3a^2H\dot{\phi} +  \Big(3a^2H^2 + k^2\Big) \phi \Big]  j_{\ell}(kr) \,.
\ea The usual flat FLRW correlation function may then be obtained using the following identity:
\ba
\sum_{\ell=0} (2\ell+1)P_{\ell}(\cos{\delta\theta})j_{\ell}(kr_1)j_{\ell}(kr_2) = \frac{\sin{ks}}{ks} \,,
\ea where $s\equiv s(z_1,z_2,\delta\theta)=\sqrt{r_1^2 + r_2^2 - 2r_1r_2\cos{\delta\theta}}$.

In LTB, the real-space radial and transverse BAO scales are  different, and are given by the peaks in the radial and transverse correlation functions. These we define as:
 \ba
 \label{xi_rad}
\mbox{radial 2PCF:} &&  \xiDelpar(z,\delta z)=\xi(z_1,z_2,0)= \sum_{\ell} (2\ell+1) \mathcal{C}_{\ell}^{\parallel}(z,\delta z),~~\delta\theta=0,~~z=z_1,~\delta{z}\equiv z_2-z_1\,, \label{xi_Delta_LTB_transverse}\\
 \label{xi_trans}
\mbox{transverse 2PCF:} && \xiDelperp(z,\delta\theta) =\xi(z,z,\delta\theta)= \sum_{\ell} (2\ell+1) P_{\ell}(\cos{\delta\theta})
\mathcal{C}_{\ell}^{\perp}(z),
\label{xi_LTB_trans}
 \ea
where the radial and transverse coefficients follow from \eqref{2pcf}:  
 \ba
  \label{c_ell_rad}
\mathcal{C}_{\ell}^{\parallel}(z,\delta z) &=& \left[(4\pi G{\apar}_1{\apar}_2)^2\rho_1\rho_2\right]^{-1} \int \frac{dk}{k} \, \, \mathcal{J}_{\ell}(z,k) \mathcal{J}_{\ell}(z+\delta z,k)\mathcal{P}_{\Phi}(k),\\
 \label{c_ell_trans}
\mathcal{C}_{\ell}^{\perp}(z) &=& \left(4\pi G\apar^2\rho\right)^{-2}\int \frac{dk}{k} \, \, {\mathcal{J}_{\ell}}^2(z,k) \mathcal{P}_{\Phi}(k).
\ea
Equations \eqref{xi_LTB}, \eqref{curlyj_LTB_1} and \eqref{xi_rad}--\eqref{c_ell_trans} summarize our new results that derive the correlation function of matter density perturbations on a radially inhomogeneous  background.

%
\begin{figure}[htpb]
\centering \epsfig{file=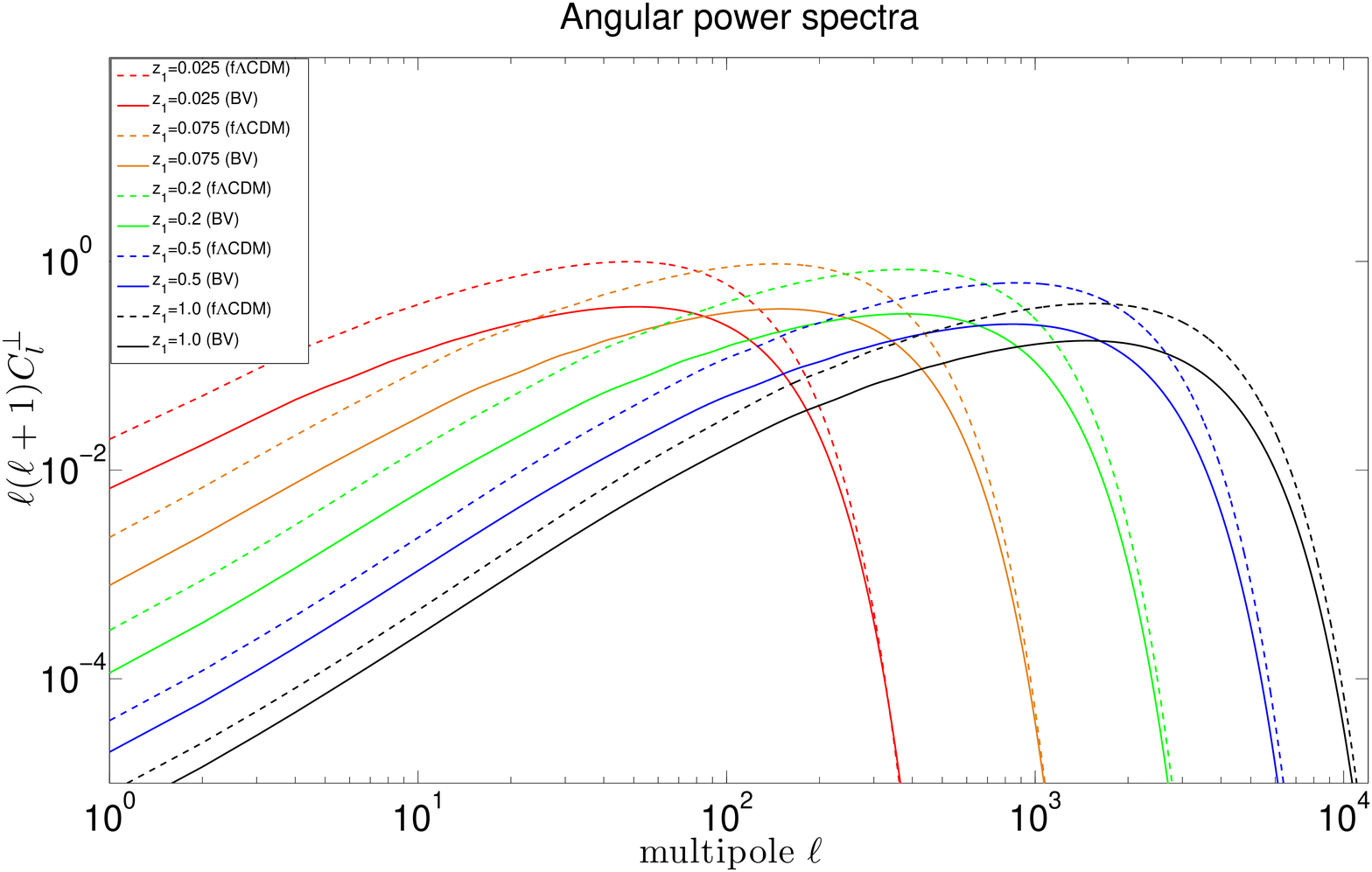,width=0.95\textwidth}
 \caption{Angular power spectra $\ell(\ell+1)\mathcal{C}_{\ell}^{\perp}$ at various redshifts for the BV and f$\LCDM$ models. (The drop in power on small scales is due to smoothing the power spectrum below 1\,Mpc.)} \label{angps}
\end{figure}

\subsection{Computation of the correlation functions and extraction of the BAO scales}
\label{sec:method}

We can now compute the correlation functions for two specific LTB models, and compare with the standard case
(see e.g.
\cite{Okumura:2007br, Padmanabhan:2008ag,Gaztanaga:2008xz, Bonvin:2011bg, Rassat:2011aa,Bertacca:2012tp}
for various approaches to compute these quantities from
galaxy surveys in the standard homogeneous framework.)
We consider the following models:
\begin{itemize}
\item Flat $\LCDM$: a concordance model, with  $\omegab h^2=0.02273$ and $\omegac h^2=0.1099$, as given by WMAP 5-year CMB-only best-fit results (see Table 6 of \cite{Hinshaw:2008kr}). Setting $h=0.7$, this implies
 \be    
\mbox{f$\Lambda$CDM:}~~~ \omegam\equiv \omegab + \omegac =0.2707,~ \omegal=1-\omegam=0.7293\,. \label{concpar}
 \ee   
 This is our benchmark model which we use to compare with void models of the type given by \eqref{void}. 

\item SV: a small void (compared to those that fit SNIa luminosity distances \cite{February:2009pv}) of type \eqref{void}, with
 \be \label{sv}
\mbox{SV:}~~~\omoin=0.2,~~\sigma=500\,\mbox{Mpc}.
 \ee
 
\item BV: a big void of type \eqref{void},  chosen so that its anisotropic expansion rates provide a good fit to observations of the average BAO scale \eqref{dzbao}. We performed a $\chi^2$ fit to measurements of $d_z$ (see Table 3 of  \cite{Blake:2011en}), and found the following best-fit parameters:
 \be \label{bv}
\mbox{BV:}~~~\omoin=0.32, ~\sigma=4.84\,\mbox{Gpc}. 
 \ee  
   
\end{itemize}
%
\begin{figure}[htpb]
\centering \epsfig{file=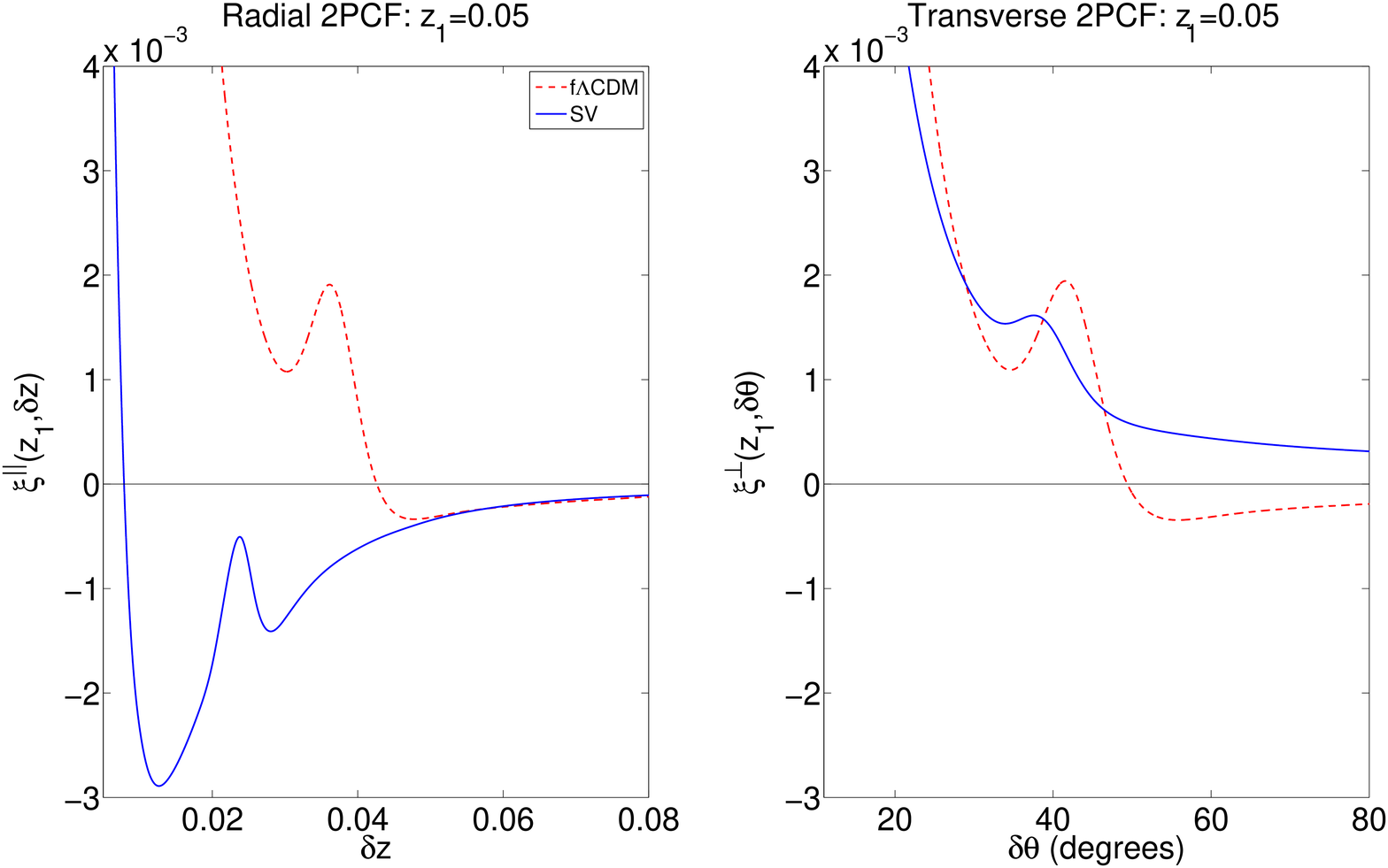,width=0.85\textwidth}
 \caption{Radial (left) and transverse (right) correlation functions for the SV and f$\LCDM$ models at $z_1=0.05$. }\label{svpcf}
\end{figure}

In the two void models, we choose FLRW initial conditions to ensure that the effects we find arise from the evolution of structure on the inhomogeneous background. We take the early-time parameters $f_b$ and $\eta$ in \eqref{ds} to be those derived from the same WMAP 5-year values used for the f$\Lambda$CDM model. This fixes the initial proper BAO scale to be the same in all models. 
The background density $\Omega_m$ and expansion rates $H_\parallel, H_\perp$ are shown for these 3 models in Fig. \ref{bgscales} (upper panels). We also show (lower panels) the geometric approximations to the radial and transverse scales, $L^\text{geo}_\|$ and $L^\text{geo}_\perp$, and the average BAO scale $d_z$  calculated from them.

Figure~\ref{fig:phi} shows the current profile and the redshift evolution of the gravitational potential  for the 3 models. Note the greater decay in the amplitude of $\phi$ for the void models, due to the presence of curvature, which explains the decrease in the overall amount of clustering relative to $\LCDM$. The normalized density perturbation is illustrated in Fig.~\ref{fig:delta}  for the 3 models. For small-scale modes (large~$\ell$), $\Delta$ scales approximately as $(1+z)^{-1}$. For the large-scale mode $\ell=2$, the `decaying' behaviour at high redshift is due to the mode entering the Hubble-scale at low redshift.

\begin{figure}[htpb]
\centering \epsfig{file=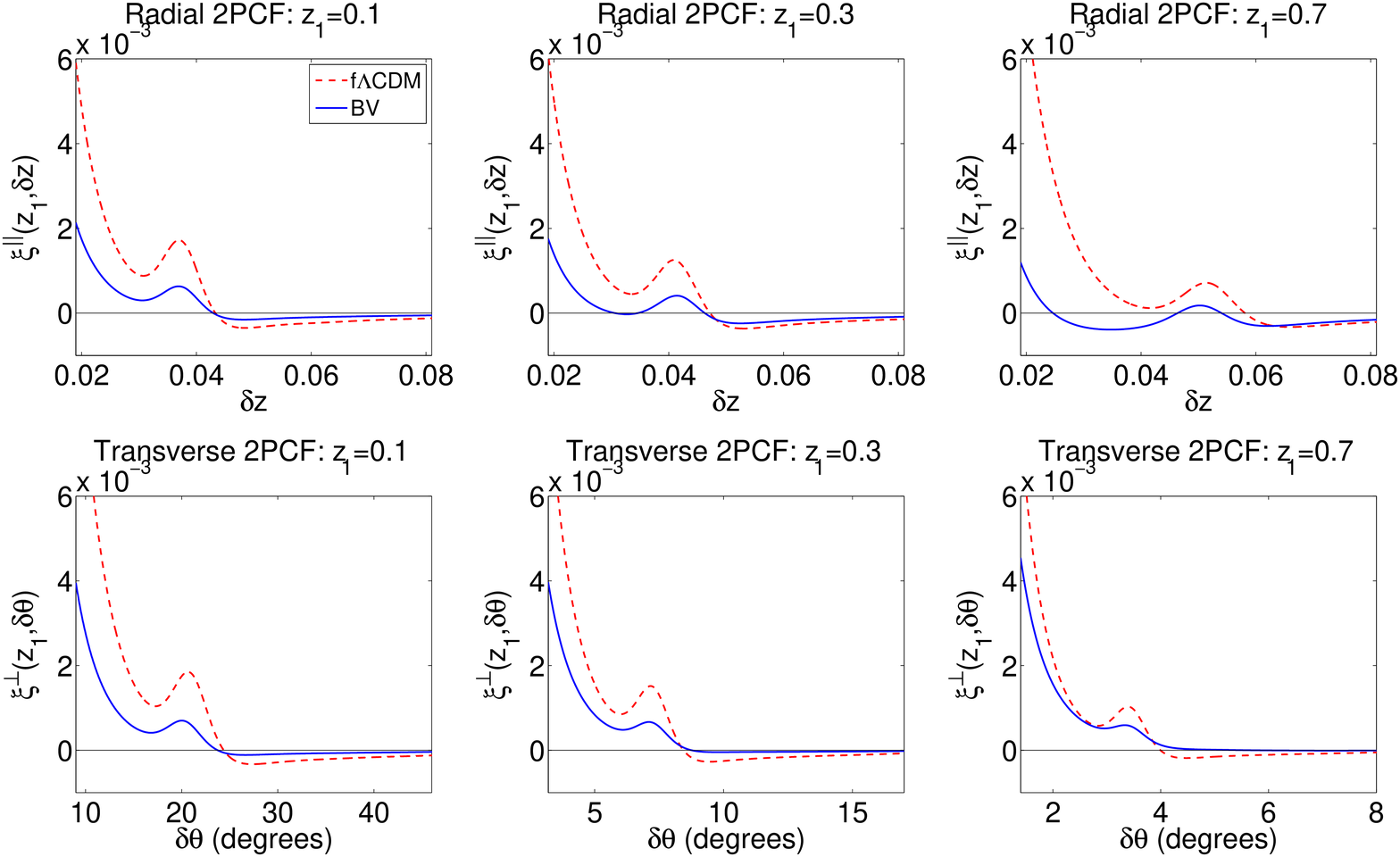,width=1.05\textwidth}
 \caption{Radial (upper) and transverse (lower) correlation functions at various redshifts for the BV and f$\LCDM$ models.  }\label{bvpcf}
\end{figure}
%
 
We calculate the correlation functions by smoothing away power on scales below 1\,Mpc, via $\mathcal{P}_{\Phi}(k) \to \mathcal{P}_{\Phi}(k) \exp[-k^2/(1\,\text{Mpc}^{-1})^2]$. This makes the sums over $\ell$ in the correlation functions \eqref{xi_Delta_LTB_transverse}, \eqref{xi_LTB_trans} converge relatively quickly (typically we require $\ell_{\rm max}(z) \lesssim 10\, r(z)/$Mpc), but without altering the resulting correlation function. Figure~\ref{angps} shows the angular power spectrum $\ell(\ell+1)\mathcal{C}_{\ell}^{\perp}$ for the BV void model compared to the concordance model. The drop in power for high $\ell$ results from the smoothing.

The correlation functions \eqref{xi_Delta_LTB_transverse}, \eqref{xi_LTB_trans} for the two void models are shown in Figs.~\ref{svpcf} and~\ref{bvpcf}. The radial correlation function $\xi^\|$, starting at various redshifts $z_1$ and extending to $z_2=z_1+\delta z$, shows the correlation of structure along a line of sight, as the observer looks into higher density regions. 
The redshift extent of the radial BAO feature is $\delta z_\text{peak}$, which is given by the location of the bump in $\xi^\|$.
The transverse correlation function $\xi^\perp$ describes the correlation across the sky in a sphere at redshift $z_1$. The angular size of the BAO is $\delta\theta_\text{peak}$, given by the bump in $\xi^\perp$. 

It is apparent from Fig.~\ref{svpcf} that for the SV model, the radial correlation function is very different from the concordance one. This is due to the large curvature gradients at low redshift, compared to void models that fit SN1a data. We neglect redshift space distortions -- but curiously, the effect of the void is qualitatively similar to the effect of redshift space distortions in FLRW (see  \cite{Montanari:2012me}). 

\begin{figure}[htpb]
\centering \epsfig{file=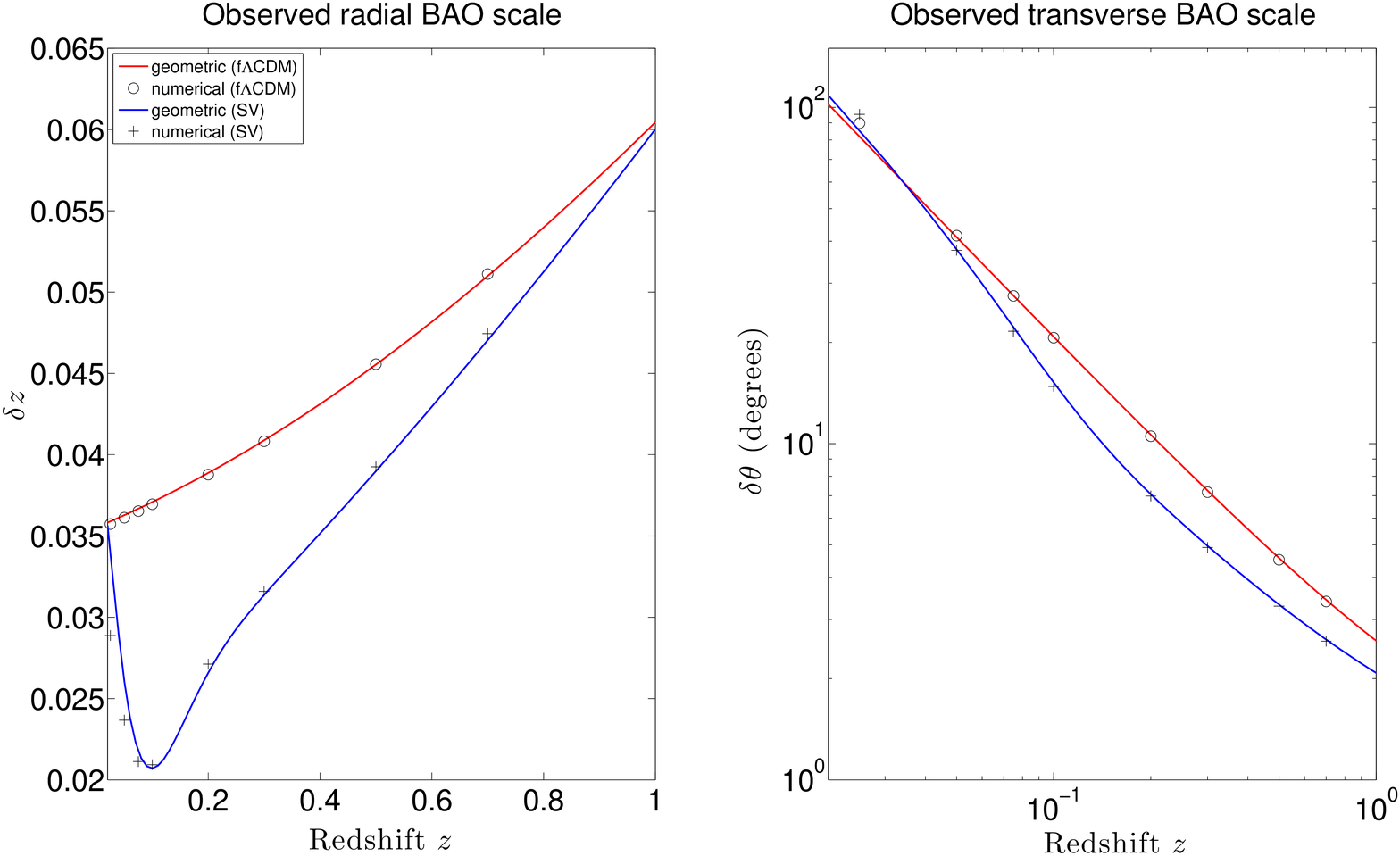,width=\textwidth}
 \caption{\textit{Upper:} Redshift  extent (left) and angular size (right) of the BAO feature for the SV and f$\LCDM$ models. \textit{Lower:} Ratio of the BAO length scales. In all plots, we show the results of the full calculation based on the correlation functions (numerical) and of the simplified geometric approximation.}\label{svbao}
\end{figure}

\begin{figure}[htpb]
\centering \epsfig{file=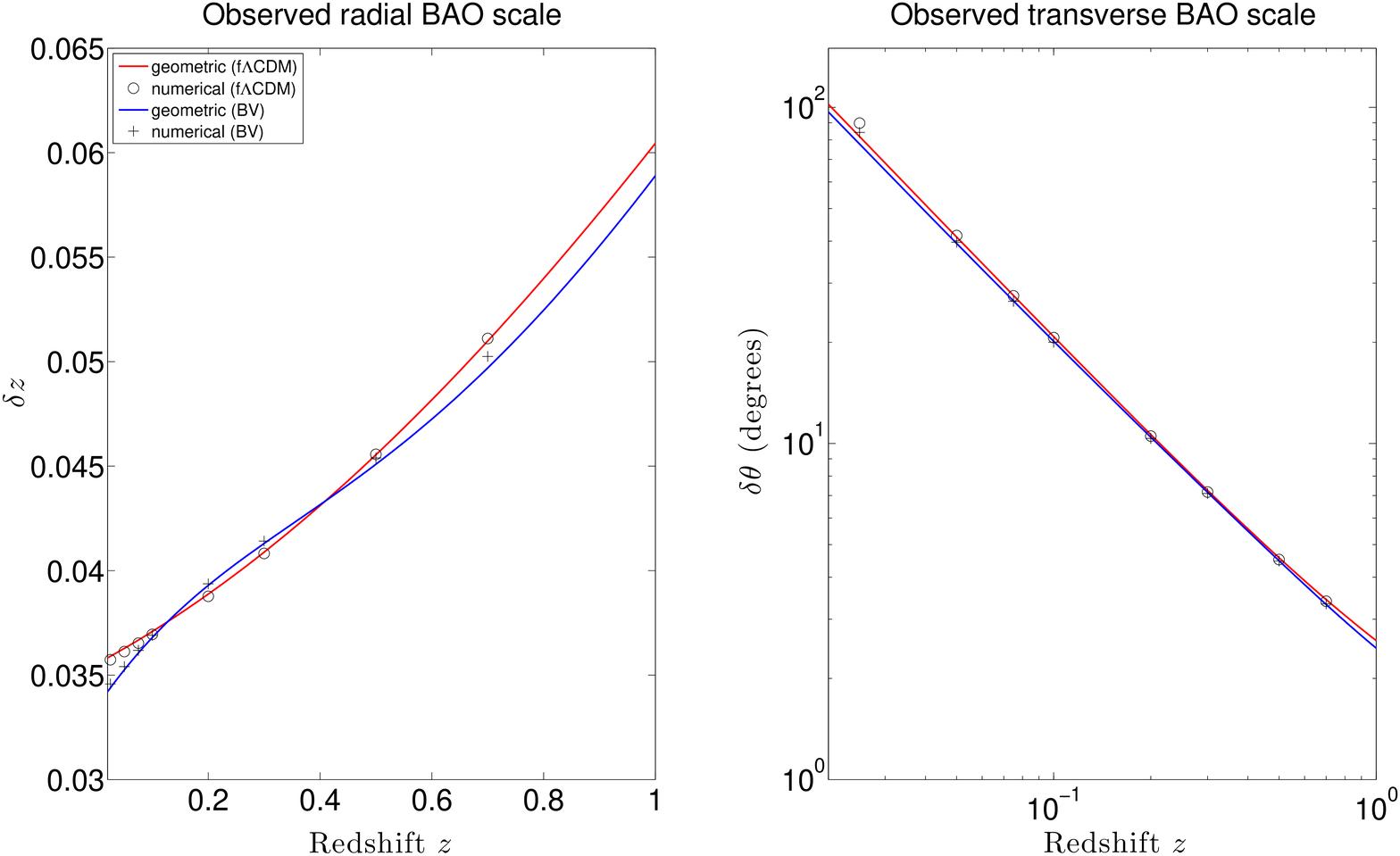,width=\textwidth}
 \caption{As in Fig. \ref{svbao}, for the BV  model. }\label{bvbao}
\end{figure}

We determined $\delta{z}_{\rm peak}$ and $\delta\theta_{\rm peak}$ numerically from the local maxima in the correlation functions. The results are shown in Figs.~\ref{svbao} and \ref{bvbao}. In these figures we also show the geometric approximations (i.e., without incorporating the effect of perturbations), 
 \be
 \label{geo}
\delta z^\text{geo}= L^\text{geo}_\|(1+z) H_\|,~~~ \delta \theta^\text{geo}= {L^\text{geo}_\perp \over d_A},
 \ee 
where $L^\text{geo}_\|,L^\text{geo}_\perp$ are given by \eqref{lexppar}.
Our results show that the geometric formulas commonly used for constraining LTB with BAO fail at the percent level. While current data are not able to resolve such differences, this may be possible with future surveys such as SKA and Euclid. Furthermore, note that the size of these corrections are of a similar order to the corrections from redshift space distortions in FLRW \cite{Montanari:2012me}. Note that the geometric formulas in \eqref{geo} give the correct observed scales for f$\LCDM$ -- except for large $\delta\theta$, for which the small-angle formula in \eqref{geo} breaks down.


\section{Conclusions}
\label{sec:discuss}

We have derived for the first time the anisotropic real-space two-point correlation function for the gauge-invariant matter density perturbation, in an LTB universe with radial inhomogeneity in the background -- summarized in \eqref{xi_LTB}, \eqref{curlyj_LTB_1} and \eqref{xi_rad}--\eqref{c_ell_trans}. For this we neglected the coupling of scalar modes with vector and tensor modes~-- which should be a good approximation, at least on the large scales relevant for the BAO. An analysis of the effects of mode-coupling, which would entail the integration of partial differential equations, is currently underway \cite{full_system_check}. We also neglected bias and redshift space distortions, since our primary focus was a comparison with the concordance model, not to test void models against data. Redshift space distortions in LTB void models deserve further investigation, in particular to check whether the FLRW formula provides a useful approximation. 

We  computed the radial and angular correlation functions for two void models, one relatively small (SV) and one Hubble-sized void that fits the average BAO data (BV) -- see Figs.~\ref{svpcf} and~\ref{bvpcf}. We used the peaks of the computed correlation functions to extract the radial and transverse BAO scales. The results were compared with the geometric approximation that has been used in all previous work, showing that the geometric approximation to the BAO scales in LTB fails at the percent level -- see Figs. \ref{svbao} and \ref{bvbao}. Future large-volume surveys, such as SKA and Euclid, may thus be able to rule out the void models on the basis of their BAO scales.

However, even if void models can be fine-tuned to reproduce the radial and transverse BAO scales, these scales represent only one feature in the galaxy correlation functions.
The void correlation functions differ significantly from those of the concordance model (Figs. \ref{svpcf} and \ref{bvpcf}). In particular, the void radial correlation can become negative (anti-correlation) before, and even at, the BAO peak, where the concordance correlation is positive. The void transverse correlation may be positive for all scales, unlike the concordance one. These features resemble the effect of redshift space distortions in FLRW (see Figs. 4 and 6 in \cite{Montanari:2012me}), since the anisotropic expansion rate in LTB can mimic the effect of radial peculiar velocities in FLRW. However,  there are significant further differences between the two models which arise from the effect of LTB perturbations.

This leads to our key final result: even if the radial and transverse BAO scales match observations, the radial and transverse {\em correlation functions} contain direct signatures of the anisotropic growth of perturbations in a non-FLRW model. These correlation functions can thus be used as direct tests of the Copernican Principle.

\acknowledgments

We thank Bruce Bassett, Marco Regis and Miguel Zumalac\'arregui for useful discussions. 
SF and RM were funded by the South African Square Kilometre Array
(SKA) Project. CC and RM were supported by the National Research Foundation
(South Africa). RM was supported by the Science \& Technology Facilities Council (UK) (grant no.
ST/H002774/1). All authors were supported by a Royal Society
(UK)/ NRF (SA) exchange grant. Computations were performed using facilities provided by the 
University of Cape Town's ICTS High Performance Computing team.


\begin{thebibliography}{99}

\bibitem{Clarkson:2010uz}
  C.~Clarkson and R.~Maartens,
  Class.\ Quant.\ Grav.\  {\bf 27}, 124008 (2010)
  [arXiv:1005.2165 [astro-ph.CO]].

\bibitem{Clarkson:2012bg}
  C.~Clarkson,
  arXiv:1204.5505 [astro-ph.CO].

\bibitem{Clifton:2011jh}
  T.~Clifton, P.~G.~Ferreira, A.~Padilla and C.~Skordis,
  Phys.\ Rept.\  {\bf 513}, 1 (2012)
  [arXiv:1106.2476 [astro-ph.CO]].
  
\bibitem{Clarkson:2011zq}
  C.~Clarkson, G.~Ellis, J.~Larena and O.~Umeh,
  Rept.\ Prog.\ Phys.\  {\bf 74}, 112901 (2011)
  [arXiv:1109.2314 [astro-ph.CO]].

\bibitem{Marra:2011ct}
  V.~Marra and A.~Notari,
  Class.\ Quant.\ Grav.\  {\bf 28} (2011) 164004
  [arXiv:1102.1015 [astro-ph.CO]].

\bibitem{Biswas:2010xm}
  T.~Biswas, A.~Notari and W.~Valkenburg,
  JCAP {\bf 1011} (2010) 030
  [arXiv:1007.3065 [astro-ph.CO]].

\bibitem{Clarkson:2010ej}
  C.~Clarkson and M.~Regis,
  JCAP {\bf 1102} (2011) 013
  [arXiv:1007.3443 [astro-ph.CO]].

\bibitem{Nadathur:2010zm}
  S.~Nadathur and S.~Sarkar,
  Phys.\ Rev.\ D {\bf 83} (2011) 063506
  [arXiv:1012.3460 [astro-ph.CO]].
  
\bibitem{GarciaBellido:2008nz}
  J.~Garcia-Bellido and T.~Haugboelle,
  JCAP {\bf 0804}, 003 (2008)
  [arXiv:0802.1523 [astro-ph]].

\bibitem{Zibin:2008vk}
  J.~P.~Zibin, A.~Moss, D.~Scott,
  Phys.\ Rev.\ Lett.\  {\bf 101}, 251303 (2008).
  [arXiv:0809.3761 [astro-ph]].

\bibitem{GarciaBellido:2008yq}
  J.~Garcia-Bellido, T.~Haugboelle,
  JCAP {\bf 0909}, 028 (2009).
  [arXiv:0810.4939 [astro-ph]].

\bibitem{Zumalacarregui:2012pq}
  M.~Zumalacarregui, J.~Garcia-Bellido and P.~Ruiz-Lapuente,
  arXiv:1201.2790 [astro-ph.CO].

\bibitem{Clarkson:2007yp}
  C.~Clarkson,
  Phys.\ Rev.\ D {\bf 76} (2007) 104034
  [arXiv:0708.1398 [gr-qc]].

\bibitem{Zibin:2008vj}
  J.~P.~Zibin,
  Phys.\ Rev.\  D {\bf 78}, 043504 (2008)
  [arXiv:0804.1787 [astro-ph]].

\bibitem{Dunsby:2010ts}
  P.~Dunsby, N.~Goheer, B.~Osano and J.~-P.~Uzan,
  JCAP {\bf 1006}, 017 (2010)
  [arXiv:1002.2397 [astro-ph.CO]].

\bibitem{Alonso:2010zv}
  D.~Alonso, J.~Garcia-Bellido, T.~Haugbolle and J.~Vicente,
  Phys.\ Rev.\ D {\bf 82} (2010) 123530
  [arXiv:1010.3453 [astro-ph.CO]].

\bibitem{Nishikawa:2012we}
  R.~Nishikawa, C.~-M.~Yoo and K.~-i.~Nakao,
  arXiv:1202.1582 [astro-ph.CO].
 
\bibitem{Clarkson:2009sc}
  C.~Clarkson, T.~Clifton and S.~February,
  JCAP {\bf 0906}, 025 (2009)
  [arXiv:0903.5040 [astro-ph.CO]].
  
 \bibitem{full_system_check}
 S. February, J. Larena and C. Clarkson, in preparation.

\bibitem{Clarkson:2007pz}
  C.~Clarkson, B.~Bassett and T.~C.~Lu,
  Phys.\ Rev.\ Lett.\  {\bf 101}, 011301 (2008)
  [arXiv:0712.3457 [astro-ph]].
  
\bibitem{Moss:2010jx}
  A.~Moss, J.~P.~Zibin and D.~Scott,
  Phys.\ Rev.\ D {\bf 83}, 103515 (2011)
  [arXiv:1007.3725 [astro-ph.CO]].
  
\bibitem{Eisenstein:2005su}
  D.~J.~Eisenstein {\it et al.}  [SDSS Collaboration],
  Astrophys.\ J.\  {\bf 633}, 560 (2005)
  [astro-ph/0501171].
    
\bibitem{Blake:2011en}
  C.~Blake {\it et al.},
  Mon.\ Not.\ Roy.\ Astron.\ Soc.\  {\bf 418}, 1707 (2011)
  [arXiv:1108.2635 [astro-ph.CO]].
  
\bibitem{February:2009pv}
  S.~February, J.~Larena, M.~Smith and C.~Clarkson,
  Mon.\ Not.\ Roy.\ Astron.\ Soc.\  {\bf 405}, 2231 (2010)
  [arXiv:0909.1479 [astro-ph.CO]].
  
\bibitem{Eisenstein:1997ik}
  D.~J.~Eisenstein and W.~Hu,
  Astrophys.\ J.\  {\bf 496}, 605 (1998)
  [arXiv:astro-ph/9709112].

\bibitem{Okumura:2007br}
  T.~Okumura, T.~Matsubara, D.~J.~Eisenstein, I.~Kayo, C.~Hikage, A.~S.~Szalay and D.~P.~Schneider,
  Astrophys.\ J.\  {\bf 676}, 889 (2008)
  [arXiv:0711.3640 [astro-ph]].

\bibitem{Padmanabhan:2008ag}
  N.~Padmanabhan and M.~J.~.~White,
  Phys.\ Rev.\  D {\bf 77}, 123540 (2008)
  [arXiv:0804.0799 [astro-ph]].

\bibitem{Gaztanaga:2008xz}
  E.~Gaztanaga, A.~Cabre, L.~Hui,
  Mon.\ Not.\ Roy.\ Astron.\ Soc.\  {\bf 399}, 1663-1680 (2009).
  [arXiv:0807.3551 [astro-ph]].

\bibitem{Bonvin:2011bg}
  C.~Bonvin and R.~Durrer,
  arXiv:1105.5280 [astro-ph.CO].

\bibitem{Rassat:2011aa}
  A.~Rassat and A.~Refregier,
  arXiv:1112.3100 [astro-ph.CO].

\bibitem{Bertacca:2012tp}
  D.~Bertacca, R.~Maartens, A.~Raccanelli and C.~Clarkson,
  JCAP, in press (2012) [arXiv:1205.5221 [astro-ph.CO]].
  
\bibitem{Hinshaw:2008kr}
  G.~Hinshaw {\it et al.}  [WMAP Collaboration],
  Astrophys.\ J.\ Suppl.\  {\bf 180}, 225 (2009)
  [arXiv:0803.0732 [astro-ph]].
  
\bibitem{Montanari:2012me}
  F.~Montanari and R.~Durrer,
  arXiv:1206.3545 [astro-ph.CO].



\end{thebibliography}
\end{document}